\documentclass{article}

\usepackage[final]{neurips_2024} 

\usepackage[utf8]{inputenc} 
\usepackage[T1]{fontenc}    
\usepackage{hyperref}       
\usepackage{url}            
\usepackage{booktabs}       
\usepackage{amsfonts}       
\usepackage{caption}
\usepackage{nicefrac}       
\usepackage{microtype}      
\usepackage{xcolor}         
\usepackage{float}
\usepackage{amsmath}
\usepackage{multirow}

\title{G-RAG: Knowledge Expansion in Material Science}

\author{
  Radeen Mostafa$^{1}$\thanks{Equal contribution.} \\
  \texttt{radeensust@gmail.com} \\
  \And
  Mirza Nihal Baig$^{2}$\footnotemark[1]\\
  \texttt{nihalmd1@gmail.com} \\
  \AND
  Mashaekh Tausif Ehsan$^{3}$\footnotemark[1]\\
  \texttt{mashaekh.tausif@gmail.com} \\
  \And
  Jakir Hasan$^{4}$\footnotemark[1] \\
  \texttt{jakir57@student.sust.edu} \\
  $^1$magicmind.me, USA\\
  $^2$Intelsense AI Limited, Dhaka, Bangladesh \\
  $^3$Department of Mechanical Engineering, Bangladesh University of Engineering and Technology \\
  $^4$Department of Computer Science and Engineering, Shahjalal University of Science and Technology \\
}

\usepackage{graphicx}
\usepackage{comment}

\begin{document}
\maketitle

\begin{abstract}
In the field of Material Science, effective information retrieval systems are essential for facilitating research. Traditional Retrieval-Augmented Generation (RAG) approaches in Large Language Models (LLMs) often encounter challenges such as outdated information, hallucinations, limited interpretability due to context constraints, and inaccurate retrieval. To address these issues, Graph RAG integrates graph databases to enhance the retrieval process. Our proposed method processes Material Science documents by extracting key entities (referred to as MatIDs) from sentences, which are then utilized to query external Wikipedia knowledge bases (KBs) for additional relevant information. We implement an agent-based parsing technique to achieve a more detailed representation of the documents. Our improved version of Graph RAG called G-RAG further leverages a graph database to capture relationships between these entities, improving both retrieval accuracy and contextual understanding. This enhanced approach demonstrates significant improvements in performance for domains that require precise information retrieval, such as Material Science. The code is available at \href{https://github.com/RadeenXALNW/G-RAG_1.0}{https://github.com/RadeenXALNW/G-RAG\_1.0}.

\end{abstract}

\section{Introduction}
LLMs exhibit impressive capabilities but encounter challenges such as hallucinations, outdated information, and untraceable, opaque reasoning. The RAG approach addresses these issues by combining the strengths of LLMs with the vast, continuously updated resources of external databases \cite{gaoa2023retrieval}. Graph-enhanced RAG methods build on this by leveraging rich semantic interconnections and relational data, enabling more precise entity linking, enhanced semantic context, and improved knowledge extraction \cite{sharma2024retrieval, ma2024think}. Additionally, researchers have introduced innovative graph-based context adaptation techniques that refine word embeddings to better capture semantic relationships, consistently outperforming traditional methods in various Natural Language Processing (NLP) tasks \cite{sandhu2024exploration, edge2024local}. Graph-based RAG provides a more nuanced and accurate representation of complex domains, enabling LLMs to generate responses with enhanced factual precision and contextual relevance \cite{park2024leveraging}. This capability is especially valuable for domain-specific applications in fields such as material science and biomedicine, where accurate and detailed information is crucial \cite{buehler2024generative, delile2024graph, peng2023knowledge}. Serving as a domain-specific knowledge server, the Semantic Context Enhancer extracts and delivers detailed descriptions of relevant concepts and entities, including their interrelationships, thereby equipping the LLM with a deeper semantic understanding \cite{peng2023knowledge}. Additionally, leveraging graph structures to improve knowledge retrieval and response generation, as exemplified by methods like AriGraph, has shown significant enhancements in decision-making and planning capabilities \cite{anokhin2024arigraph}. This study explores the improvement of information retrieval and knowledge generation in complex, specialized domains through the integration of the G-RAG pipeline, addressing limitations of existing approaches and advancing performance in targeted fields.

\setcounter{footnote}{0}
\section{Methodology}
The retrieval process of Naive RAG includes a diverse range of MatIDs, which ensures variety but can also introduce less relevant information. This issue can be mitigated through prompt engineering in the RAG configuration, allowing the LLM to continue generating accurate responses \cite{merth2024superposition}. However, there are two main limitations to this approach. First, LLMs have a fixed context window, which restricts the number of tokens they can process simultaneously. This limitation hinders the model's ability to manage large volumes of retrieved data effectively \cite{chen2023extending}, especially when the dataset is extensive and varied. Despite advancements like Google’s Gemini, which uses a caching system to handle extended contexts, the fixed context window of LLMs remains a significant constraint \cite{hu2024memserve, wang2024searching}. Although providing the model with more relevant information might seem beneficial, increasing the context length does not necessarily improve the accuracy of information retrieval or response generation \cite{liu2024lost}. 
This problem becomes even more pronounced when the retrieved context includes a mix of diverse but only marginally relevant data, potentially diluting the focus on the critical entities or concepts needed for an accurate response\footnote{\href{https://towardsdatascience.com/building-a-biomedical-entity-linker-with-llms-d385cb85c15a}{https://towardsdatascience.com/building-a-biomedical-entity-linker-with-llms-d385cb85c15a}}. This is where Graph RAG proves to be valuable, as it enhances the retrieval process by focusing on the most relevant information.

\subsection{Graph RAG vs G-RAG}
\begin{figure}[!htpb]
    \centering
    \includegraphics[width=0.72\textwidth]{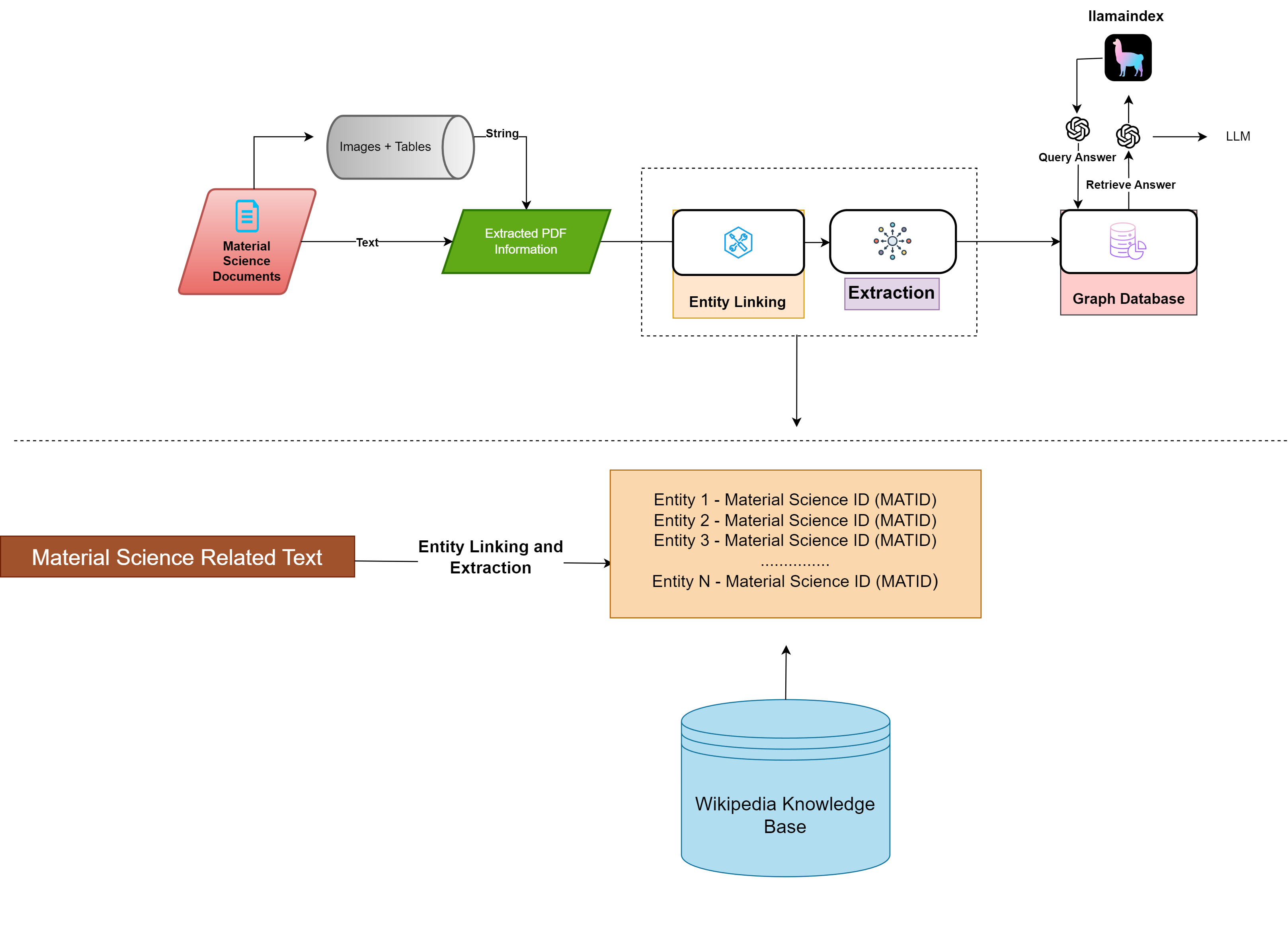}
    \caption{Architecture of G-RAG System}
    \label{fig:g_rag_pipeline}
\end{figure}

Graph RAG effectively merges the strengths of retrieval-based and generative methods to enhance LLMs' capability to generate accurate, relevant, and contextually enriched responses \cite{peng2024graph}. While supplying an LLM with text chunks from extensive documents may result in issues with context, factual precision, and language coherence, Graph RAG addresses these limitations by utilizing a knowledge graph as a source of structured, factual information \cite{edge2024local}. The knowledge graph provides detailed entity information, including attributes and relationships, allowing the LLM to gain a deeper understanding and produce more informed, precise responses. In our G-RAG system, entity linking is a fundamental component, enabling the extraction of specific entities (key terms or concepts) from the text using an entity extractor like a Span Parser. These identified entities are then used to query an external retriever, which fetches relevant MatIDs and their corresponding information from a Wikipedia knowledge base\footnote{\href{https://huggingface.co/relik-ie/relik-reader-deberta-v3-large-re-wikipedia}{https://huggingface.co/relik-ie/relik-reader-deberta-v3-large-re-wikipedia}}. This targeted retrieval process ensures that the selected MatIDs are highly relevant and accurate, thereby preserving the integrity and relevancy of the constructed knowledge graph\footnote{\href{https://neo4j.com/developer-blog/entity-linking-relationship-extraction-relik-llamaindex/}{https://neo4j.com/developer-blog/entity-linking-relationship-extraction-relik-llamaindex/}}. Following this, an LLM formulates a query that is sent to the graph database. The graph database retrieves relevant information, which is processed by the LLM to generate a final, comprehensive response. We set a limit on the number of nodes retrieved to ensure the data fits within the context length of the LLM model \ref{node retrieval}. The complete architecture of our G-RAG system is illustrated in Figure \ref{fig:g_rag_pipeline}.

\subsection{PDF Parsing}
\label{headings}
We parse PDFs by categorizing their content into text, figures, and tables. For figure extraction, we employ the Phi-3.5 Vision Instruct model, specifically tailored to identify material science-related images using a vision agent system. We utilize Microsoft's Table Transformer in the tabular data extraction process. 
Furthermore, we apply a smart chunking technique to enhance the precision of data segmentation. Accurate parsing is essential for subsequent tasks such as Entity Linking, Relation Extraction, and Graph Retrieval Augmented Generation, as it ensures the accuracy and relevance of the answers retrieved from the database. Appendix \ref{document_parsing} provides a detailed overview of our document parsing process.

\subsection{Entity Linking and Relation Extraction}
Entity Linking (EL) refers to the process of mapping ambiguous mentions in a text to specific, identifiable named entities within a knowledge base \cite{tedeschi2021named}. It involves recognizing all potential entities mentioned in the given input and accurately associating them with corresponding entries in a reference knowledge base, such as Wikipedia. Relation Extraction (RE) refers to the process of identifying and classifying semantic relationships between entities mentioned within a given text. This task involves mapping the detected entities to specific relation categories defined in a reference knowledge base, such as Wikipedia. The entity linking and relation extraction process is depicted in Appendix \ref{entity_linking}.

\subsection{Span Parser}
The Span Parser module functions as our G-RAG system's initial information retrieval component, employing an approach inspired by the Retrieval Process \cite{orlando2024relik}. This module operates on the principle of semantic similarity between the current knowledge base (KB) and a comprehensive collection of textual passages (Wikipedia Database) representing entities and relations. At its core, the Span Parsing module utilizes an encoder to generate dense vector representations of both the knowledge base (KB) \( q \) and each passage \( p \) in the additional knowledge base collection. These representations, denoted as \( E(q) \) and \( E(p) \) respectively, are high-dimensional embedding that capture the semantic content of the text. The module computes a similarity score between the current Knowledge Base and additional Knowledge Base (Wikipedia data) using a dot product operation, yielding the most relevant relations with respect to the extended knowledge base \( q \):
\[\text{sim}(q, p) = E(q)^\top \cdot E(p)\]
This score quantifies the relevance of each passage of the additional knowledge base to the given current KB passage's sentence, enabling the module to rank and retrieve the most pertinent information.

\subsection{Passage Processor}
The Passage Processor (PP) component in our G-RAG system employs a unified approach to process the existing knowledge base and retrieved passages. Given a current Knowledge Base (KB) \( Q \) and a set of \( N \) retrieved passages \(\{P_1, \ldots, P_n\}\), the Passage Processor constructs chunks of current KB. In each chunk, we utilize each input sequence \( S = [Q; \tau_0; P_1; \tau_1; \ldots; P_n; \tau_n] \), where \( \tau_i \) are delimiter tokens. This sequence is encoded using a Transformer model \( T \), producing contextual embedding \( E = T(S) \). The Passage Processor subsequently identifies relevant spans within \( Q \) through a two-stage process \cite{orlando2024relik}. Initially, it computes start probabilities \( P^s(q_i) \) for each token \( q_i \) in \( Q \) using a learned function \( f^s(E) \). Subsequently, for each potential start position \( s \), it calculates end probabilities \( P^e(q_j \mid s) \) for tokens \( q_j \) (where \( j \geq s \)) using another learned function \( f^e(E, s) \). This formulation enables the prediction of overlapping spans, enhancing the model's capability to handle complex queries. During the process, spans \( (s, e) \) are predicted if \( P^s(q_s) > \theta_s \) and \( P^e(q_e \mid s) > \theta_e \), where \( \theta_s \) and \( \theta_e \) are predefined thresholds. This design enables the Passage Processor to process the entire knowledge base chunk by chunk efficiently, identifying relevant text spans for downstream tasks such as entity linking and relation extraction.

\section{Experimental Settings}
Our dataset consists of ten carefully designed handwritten queries, aimed at evaluating and differentiating the capabilities of various RAG systems. Sample queries from this dataset are presented in Appendix \ref{examples}.
To evaluate the performance of RAG systems, we employ various metrics, including correctness, faithfulness, context, and answer relevancy scores. Correctness assesses the accuracy of the generated response, while faithfulness evaluates the factual accuracy based on the retrieved documents. Finally, the context and answer relevancy score measures how well the response aligns with the given query. A detailed description of these evaluation metrics is provided in Appendix \ref{metrics}.
For entity linking and relation extraction, we use the relik-entity-linking-large model\footnote{\href{https://huggingface.co/sapienzanlp/relik-entity-linking-large}{https://huggingface.co/sapienzanlp/relik-entity-linking-large}}, while the jina-embeddings-v2-base-en model\footnote{\href{https://huggingface.co/jinaai/jina-embeddings-v2-base-en}{https://huggingface.co/jinaai/jina-embeddings-v2-base-en}}, with a sequence length of 8192, is employed for embeddings. Additionally, we utilize LLama 3.1 8B and LLama 3.1 70B as large language models, both of which produce comparable results.

\section{Results and Discussion}
This section presents all of our experimental results. We conducted the computational tasks using the NVIDIA Tesla A100 Ampere 40 GB GPU. The performance of the Naive RAG, Graph RAG, and the G-RAG system was evaluated using our dataset. Appendix \ref{response} provides example queries and the corresponding responses from the RAG systems, evaluated across different metrics. The experimental results are summarized in Table \ref{tab:experimental_results}.

\begin{table}[!htbp]
\centering
\caption{Experimental Results}
\vspace{1em}
\begin{tabular}{@{}llccc@{}}
\toprule
\textbf{Pipeline} & \textbf{Score} & \textbf{No. of queries} & \textbf{Mean} & \textbf{Standard Deviation} \\
\midrule

\multirow{3}{*}{Naive RAG} & Correctness & \multirow{3}{*}{10} & 2.43 & 1.51\\
 & Faithfulness &  & 0.70 & 0.48\\
 & Relevancy &  & \textbf{0.39} &  0.28\\
 \midrule

 \multirow{3}{*}{Graph RAG} & Correctness & \multirow{3}{*}{10} &  3.30 & 2.00\\
 & Faithfulness &  & \textbf{0.90} & \textbf{0.32}\\
 & Relevancy &  & 0.18 & \textbf{0.26}\\
\midrule

\multirow{3}{*}{G-RAG} & Correctness & \multirow{3}{*}{10} & \textbf{3.90} & \textbf{1.10}\\
 & Faithfulness &  & \textbf{0.90} & \textbf{0.32}\\
 & Relevancy &  &  0.34 & 0.32\\
 
\bottomrule
\end{tabular}
\label{tab:experimental_results}
\end{table}

The comparative analysis of three RAG pipelines - Vector/Naive RAG, G-RAG, and Graph RAG - showed interesting patterns in their performance across three critical dimensions. A one-way Analysis of Variance (ANOVA) as described in Appendix \ref{anova} was performed, examining correctness $F(2,24) = 2.39$, $p = 0.113$, faithfulness $F(2,27) = 1.04$, $p = 0.368$, and context and answer relevancy $F(2,27) = 1.04$, $p = 0.368$. While no statistically significant differences were found at the standard significance level ($\alpha = 0.05$), the descriptive statistics highlighted meaningful variations in performance. Specifically, Vector/Naive RAG outperformed the others in terms of context relevancy, with a mean score of 0.3875. This was followed by G-RAG (mean score of 0.3375), while Graph RAG exhibited the lowest mean score of 0.1750. The substantial standard deviations observed across all pipelines, ranging from 0.2630 to 0.3162, suggest notable performance variability depending on the query. This variability highlights the challenge of consistency in RAG systems. The superior performance of G-RAG over the basic Graph RAG can be attributed to the inclusion of a material science knowledge base, emphasizing the critical role of domain-specific knowledge in enhancing model accuracy. The superior context relevancy performance of the traditional Vector/Naive RAG challenges the assumption that graph-based approaches inherently provide better retrieval capabilities. G-RAG has proven to be a well-rounded solution, effectively balancing the metrics of correctness, relevancy, and faithfulness. The significant drop in relevancy scores for Graph RAG highlights the critical role of entity linking in G-RAG’s design. This suggests that the effectiveness of knowledge integration mechanisms, including entity linking, plays a substantial role in improving retrieval performance. These findings indicate that while graph-based approaches show promise, their success heavily depends on the quality of knowledge integration and the sophistication of the entity-linking.

\section{Conclusion and Future Work}
Our findings indicate that integrating graph-based techniques and ensuring robust entity linking with external databases can significantly enhance the performance of the Graph RAG pipeline, particularly in terms of response relevance and accuracy. This approach also mitigates the challenge of maintaining relevance observed in standard Graph RAG implementations. Future work could include developing a larger knowledge base tailored to material science as an extended information source, as well as creating a material science-specific entity linking model. We also aim to explore this method in other domains where retrieving accurate, precise, and relevant information is essential. Additionally, establishing a comprehensive evaluation metric for Graph RAG would provide deeper insights into the process and its effectiveness.


\appendix
\newpage

\section{Appendix}

\subsection{Node Selection Process for retrieving information}
\label{node retrieval}
Let:
\begin{itemize}
    \item \( q \) be the user's question.
    \item \( D = (N, R) \) be the graph database, where \( N \) is the set of nodes and \( R \) is the set of relationships.
    \item \( \text{LLM} \) be the Large Language Model used to generate the answer.
    \item \( \text{ContextLength}_{\text{max}} \) be the maximum context length allowed by the LLM.
\end{itemize}

After that, now let's define the keyword extraction function \( K: Q \rightarrow 2^{\Sigma^*} \):

\[
K(q) = \{ \text{lemma}(t) \mid t \in \text{Tokens}(q), \ \text{POS}(t) \in S, \ |\text{lemma}(t)| > l_{\text{min}} \}
\]
where:
\begin{itemize}
    \item \( \text{Tokens}(q) \) is the set of tokens from question \( q \).
    \item \( \text{lemma}(t) \) is the lemmatized form of token \( t \).
    \item \( \text{POS}(t) \) returns the part-of-speech tag of token \( t \).
    \item \( S = \{\text{NOUN}, \text{PROPN}, \text{ADJ}, \text{VERB}\} \).
    \item \( l_{\text{min}} \) is the minimum length of a lemma.
\end{itemize}

Here, the relevance function is \( \text{Rel}: \Sigma^* \times \Sigma^* \rightarrow \{0,1\} \):

\[
\text{Rel}(s, k) =
\begin{cases}
1, & \text{if } k \subseteq \text{lower}(s) \\
0, & \text{otherwise}
\end{cases}
\]

where \( \text{lower}(s) \) converts string \( s \) to lowercase.

Now, let's define the node selection function \( N_K: 2^{\Sigma^*} \times N \rightarrow 2^N \):

\[
N_K(K, N) = \left\{ n \in N \mid \exists k \in K, \ \text{Rel}(n.\text{text}, k) = 1 \right\}
\]

And, then comes the relationship selection function \( R_K: 2^{\Sigma^*} \times R \rightarrow 2^R \):

\[
R_K(K, R) = \left\{ r \in R \mid \exists k \in K, \ \text{Rel}(r.\text{text}, k) = 1 \right\}
\]

Apply limits to ensure the context fits within the LLM's context length:

\[
|N_K| \leq N_{\text{max}}, \quad |R_K| \leq R_{\text{max}}
\]

where \( N_{\text{max}} \) and \( R_{\text{max}} \) are determined based on \( \text{ContextLength}_{\text{max}} \).

Here, the context construction function will be \( C: 2^N \times 2^R \rightarrow \Sigma^* \):

\[
C(N_K, R_K) = \text{Concat}\left( \left\{ n.\text{text} \mid n \in N_K \right\} \cup \left\{ r.\text{text} \mid r \in R_K \right\} \right)
\]

where \( \text{Concat} \) concatenates the text attributes into a single string.

After that, we can generate data using prompt generation.

Define the prompt generation function \( P: Q \times \Sigma^* \rightarrow \Sigma^* \):

\[
P(q, C) = \text{Template}(q, C)
\]

where \( \text{Template}(q, C) \) is a predefined template that incorporates the question \( q \) and context \( C \).

The final answer \( a \) is obtained by passing the prompt \( P(q, C) \) to the LLM:

\[
a = \text{LLM}(P(q, C))
\]

\subsection{Documents Parsing Method}
\label{document_parsing}

\begin{figure}[H]
    \centering
    \includegraphics[width=15cm, height=6cm, keepaspectratio]{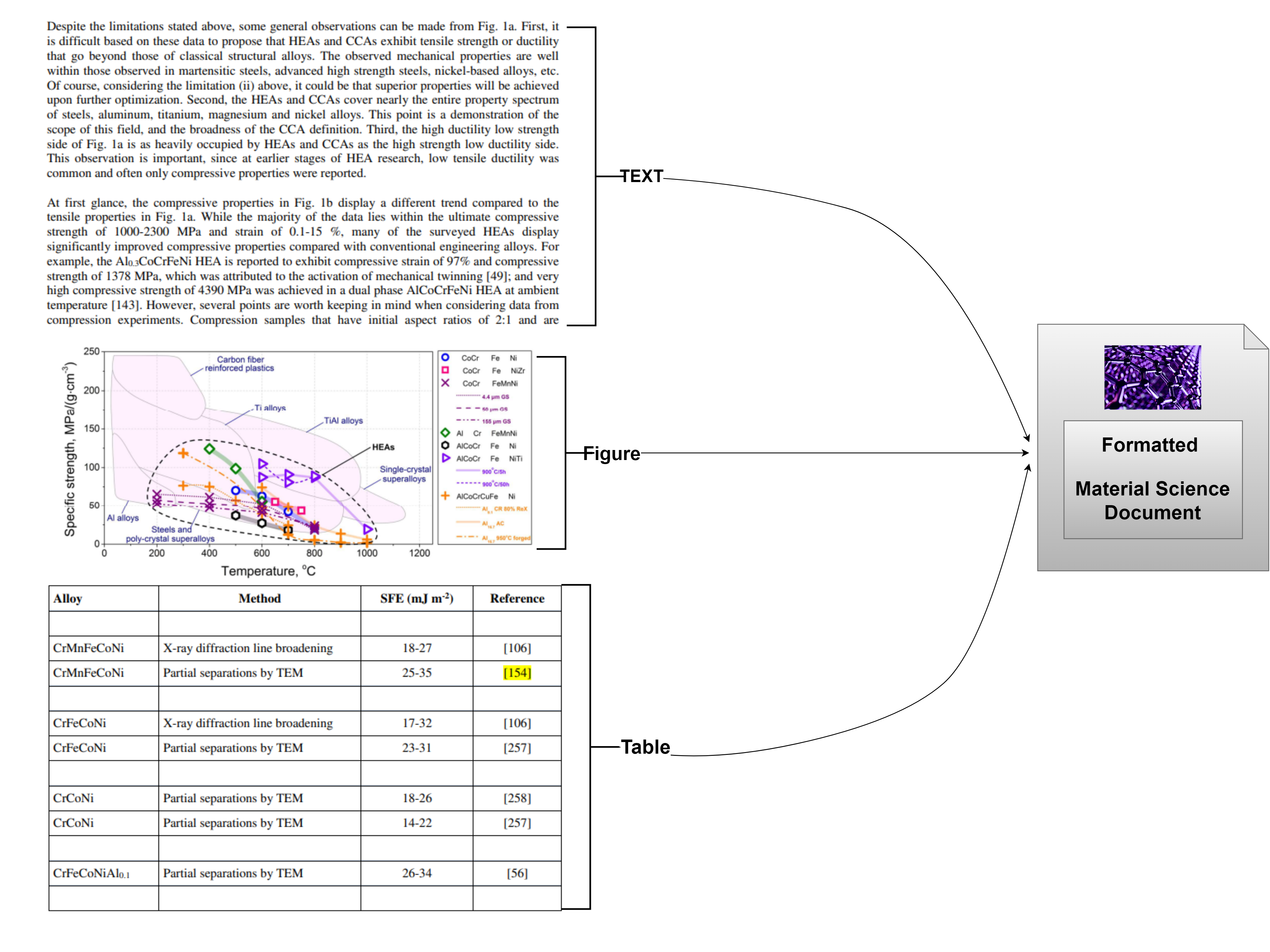}
    \caption{Document Parsing}
    \label{fig:pdf}
\end{figure}

\begin{figure}[H]
    \centering
    \includegraphics[width=15cm, height=7cm, keepaspectratio]{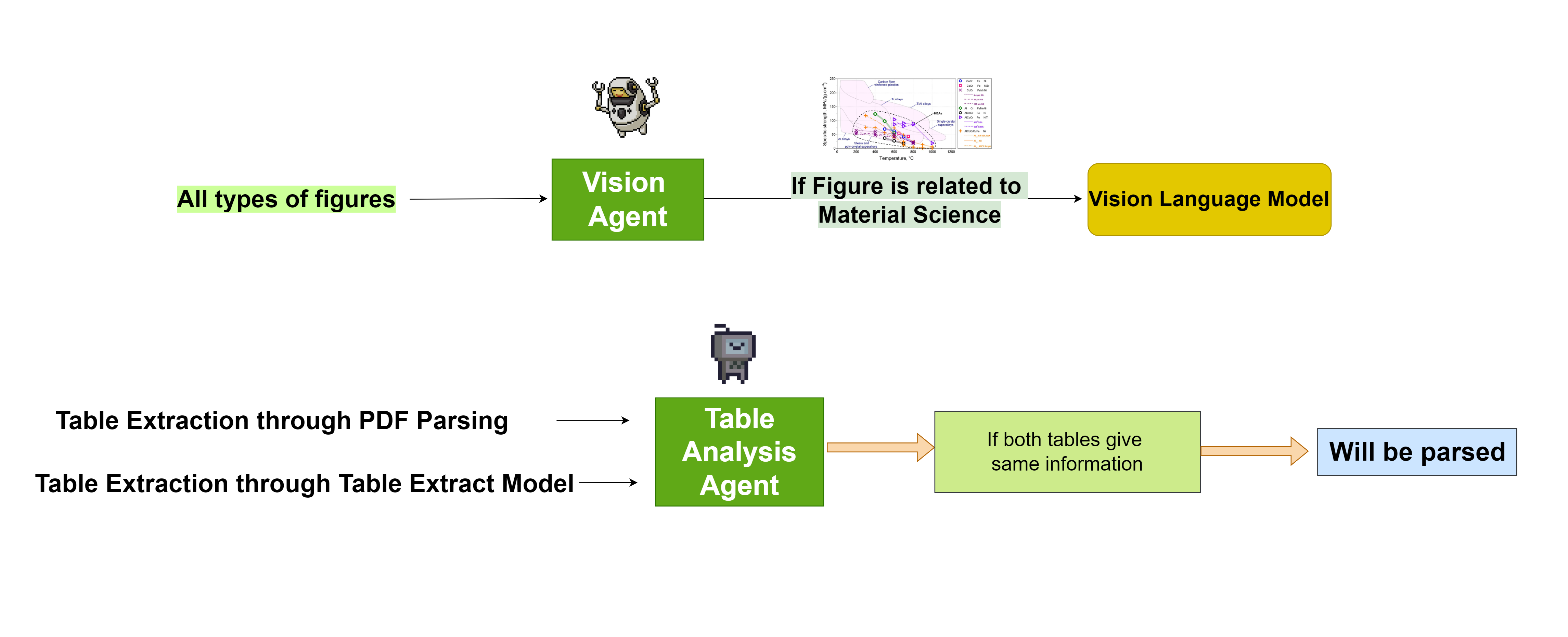}
    \caption{Validity Check by Agent System}
    \label{fig:pdf_extraction_agent}
\end{figure}

This section illustrates our document parsing pipeline, as shown in Figures \ref{fig:pdf} and \ref{fig:pdf_extraction_agent}. Efficient document parsing is crucial for enabling RAG systems to generate responses with high factual accuracy and precision.

\subsection{Entity Linking and Relation Extraction} 
\label{entity_linking}

\begin{figure}[H]
    \centering
    \includegraphics[width=35cm, height=15cm, keepaspectratio]{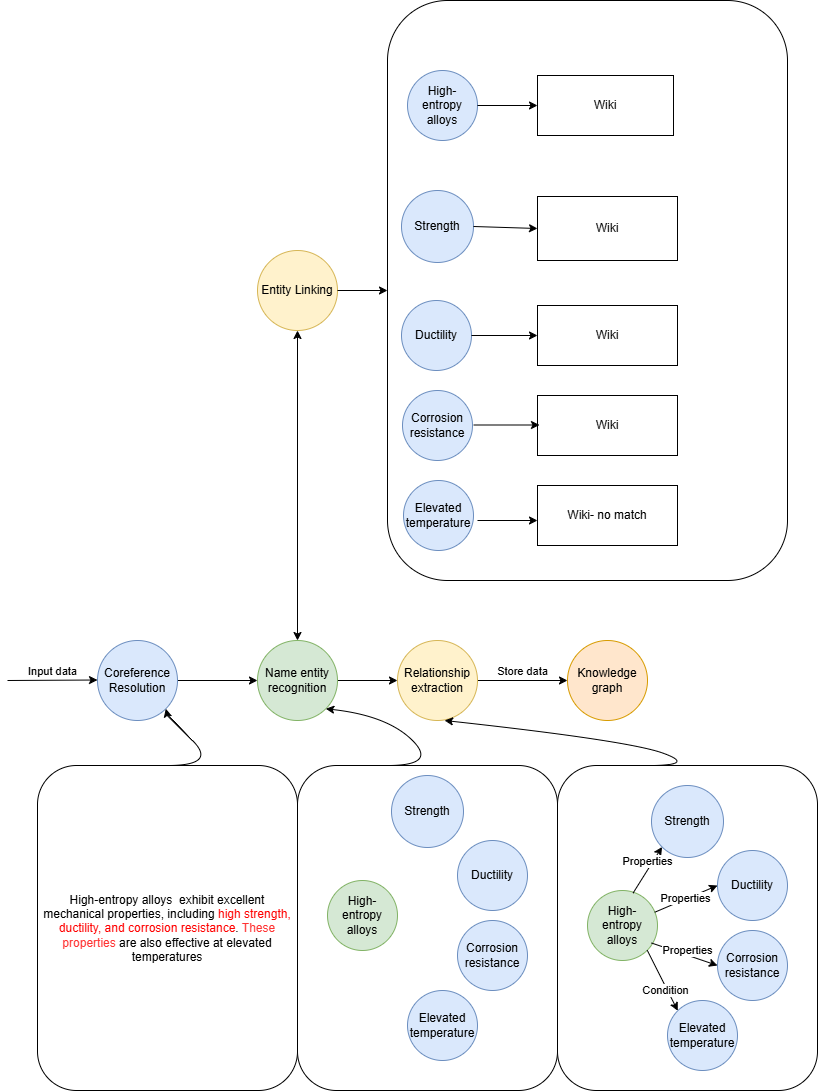}
    \caption{Entity Linking and Relation Extraction} 
    \label{fig:fig1}
\end{figure}

\begin{figure}[H]
    \centering
    \includegraphics[width=15cm, height=10cm, keepaspectratio]{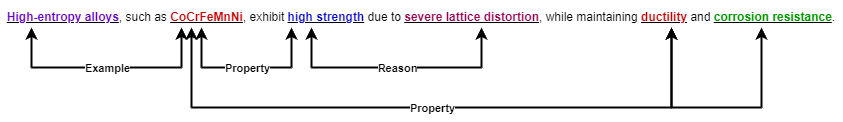}
    \caption{Entity Linking}
    \label{fig:fig2}
\end{figure}

\begin{figure}[H]
    \centering
    \includegraphics[width=55cm, height=16cm, keepaspectratio]{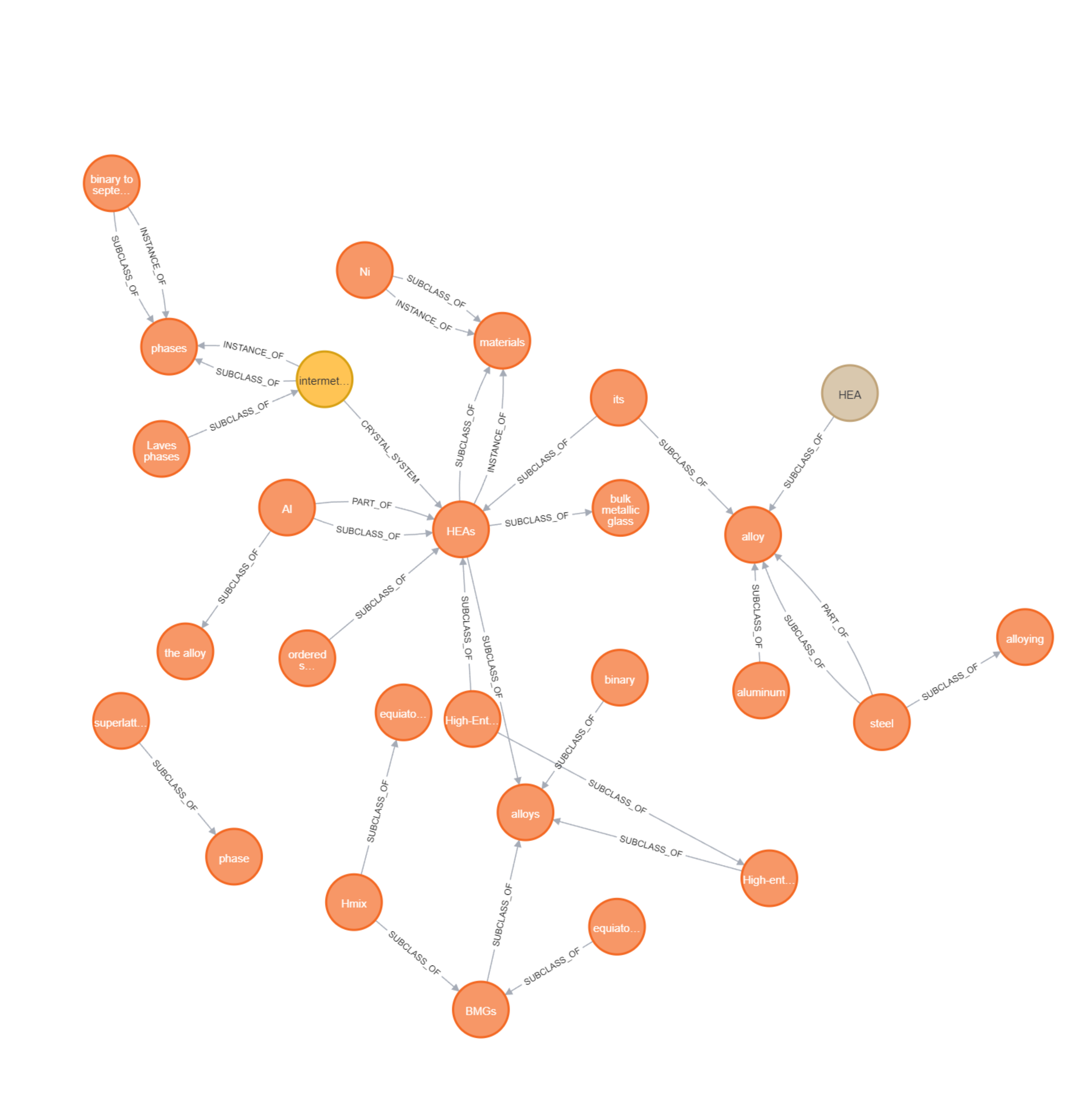}
    \caption{Relationship among Various High-entropy alloy Components} 
    \label{fig:fig3}
\end{figure}

\begin{figure}[H]
    \centering
    \includegraphics[width=20cm, height=7cm, keepaspectratio]{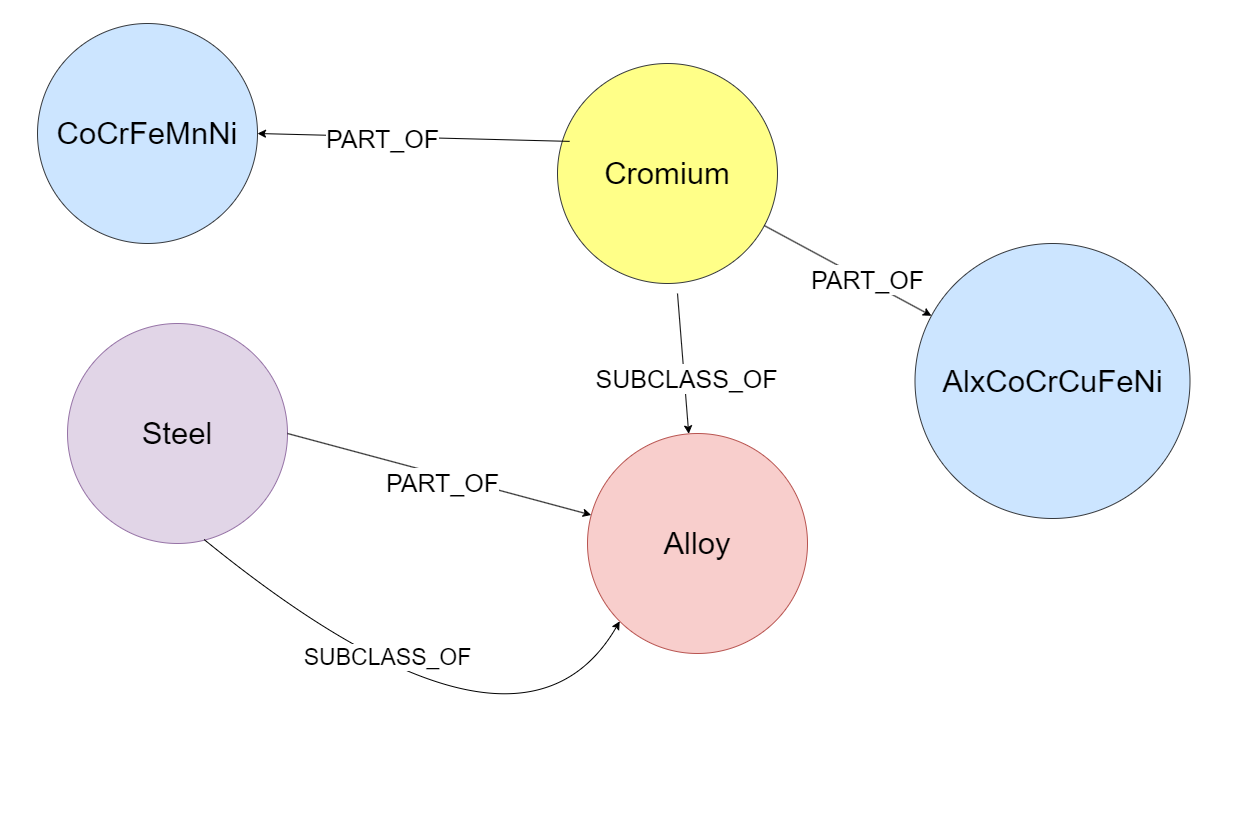}
    \caption{Another Relationship among Various High-entropy alloy Components}
    \label{fig:fig4}
\end{figure}

In this section, we provide a visual representation of the entity linking and relation extraction process, as depicted in Figures \ref{fig:fig1}, \ref{fig:fig2}, \ref{fig:fig3}, and \ref{fig:fig4}. These processes are essential components of our G-RAG system.

\textbf{Coreference Resolution:} Coreference resolution, mentioned in Figure \ref{fig:fig1} involves identifying different expressions in a text that refer to the same entity. This process is crucial for understanding the relationships between various mentions of an entity within a given context.

\subsection{Examples from Our Dataset}
\label{examples}

In this section, we present sample queries from our dataset in Table \ref{tab:example_queries}, covering a range from simple to more complex queries.

\newpage

\begin{table}[h!]
    \centering
    \caption{Example Queries}
    \renewcommand{\arraystretch}{1.75} 
    \vspace{1em}
    \begin{tabular}{p{5cm} p{8cm}}
        \toprule
        \textbf{Query} & \textbf{What is the yield strength of the CrMnFeCoNi alloy at 600 K, 700 K with 4 µm grain size?} \\
        \textbf{Ground Truth} & The yield strength of the CrMnFeCoNi alloy at \(600\) K is \(290\) MPa, and at \(700\) K, it is \(285\) MPa. \\

        \midrule
        \textbf{Query} & \textbf{What is the CRSS of CrMnFeCoNi at the tension in room temperature?} \\
        \textbf{Ground Truth} & The Critical Resolved Shear Stress (CRSS) of the CrMnFeCoNi alloy has been measured at \(53\) MPa at room temperature and \(175\) MPa at \(77\) K. \\

        \midrule
        \textbf{Query} & \textbf{What is the stacking fault energy of CrCoNi?} \\
        
        \textbf{Ground Truth} & The stacking fault energy of CrCoNi is \(18 - 26 \, \text{mJ/m}^2\).\\

        \midrule
        \textbf{Query} & \textbf{At room temperature, what is the Hall-Petch slope of the cantor alloy?} \\
        
        \textbf{Ground Truth} & At room temperature, the Hall-Petch slope of the cantor alloy was determined to be \(494 \, \text{MPa} \, \mu\text{m}^{-1/2}\).\\

        \midrule
        \textbf{Query} & \textbf{What is the stacking fault energy of the cantor alloy?} \\
        \textbf{Ground Truth} & The stacking fault energy of the cantor alloy was estimated to be \(\sim 30 \, \text{mJ} \, \text{m}^{-2}\).\\

        \midrule
        \textbf{Query} & \textbf{What is the yield strength and ultimate tensile strength of TiZrNbHfTa after 1000°C annealing?} \\
        \textbf{Ground Truth} & After \(1000\,^{\circ}\text{C}\), the yield strength will be \(1145 \, \text{MPa}\), and the ultimate tensile strength will be \(1262 \, \text{MPa}\).\\

        \midrule
        \textbf{Query} & \textbf{What is the CRSS of CrFeCoNiAl0.3 in compression at room temperature?} \\
        \textbf{Ground Truth} & CRSS of CrFeCoNiAl0.3 in compression at room temperature is \(54\) MPa.\\

        \bottomrule
    \end{tabular}
    \label{tab:example_queries}
    \renewcommand{\arraystretch}{1} 
\end{table}

\newpage

\subsection{LLM RAG Evaluation Metrics}
\label{metrics}
This section provides detailed descriptions of the various evaluation metrics used for RAG systems.

\subsubsection{Correctness}
Given a query \( q \), a generated answer \( g \), and an optional reference answer \( r \), the \texttt{CorrectnessEvaluator} computes a score \( s \) using an LLM. This score is then compared against a threshold \( T \) to determine whether the generated answer is correct or passing.

\begin{align*}
\text{Prompt} &\quad \text{Constructed from } q, g, r \\
E(g, q, r) &\quad \text{LLM Response to Prompt} \\
(s, \text{reasoning}) &\quad \text{parser\_function}(E(g, q, r)) \\
\text{passing} &\quad s \geq T \\
\text{EvaluationResult} &\quad \{q, g, \text{passing}, s, \text{reasoning}\}
\end{align*}

\subsubsection{Faithfulness Evaluation}

Given a query \( q \), a response \( r \), and a set of context documents \( C \), the \texttt{FaithfulnessEvaluator} performs the following steps:

\begin{align*}
\text{Context Documents} &\quad \text{Transform } C \text{ into } \text{Document objects} \\
\text{Index} &\quad \text{Create } \text{SummaryIndex} \text{ from Document objects} \\
\text{Query Engine} &\quad \text{Create } \text{query engine using } \text{LLM, eval\_template, and refine\_template} \\
\text{Evaluation} &\quad \text{Perform } \text{a query} \text{ on the response using the query engine} \\
\text{Raw Response} &\quad \text{Obtain } \text{raw\_response\_txt} \text{ from the query engine} \\
\text{Passing} &\quad 
\begin{cases}
\text{True} & \text{if } \text{yes} \text{ is found in } \text{raw\_response\_txt} \\
\text{False} & \text{otherwise}
\end{cases} \\
\text{Score} &\quad 
\begin{cases}
1.0 & \text{if passing is True} \\
0.0 & \text{otherwise}
\end{cases} \\
\text{Feedback} &\quad \text{raw\_response\_txt}
\end{align*}

The evaluation result is given by:
\[
\text{EvaluationResult} = \{ q, r, C, \text{passing}, \text{score}, \text{feedback} \}
\]

\subsubsection{Answer Relevancy}
\textit{Let} \( q \) be the query, \( r \) the response, and \( \{c_1, c_2, \dots, c_n\} \) the contexts.
\textit{Define the following:}
\[
\text{Documents} = \{d_i \mid d_i = \text{Document}(\text{text} = c_i) \text{ for } i = 1, 2, \dots, n\}
\]
\[
\text{Index} = \text{SummaryIndex}(\text{Documents})
\]
\[
\text{query\_response} = \text{Question: } q \text{ Response: } r
\]

\textit{Evaluate the query-response pair with:}

\[
\text{response\_obj} = \text{QueryEngine}(\text{Index}).\text{aquery}(\text{query\_response})
\]
\textit{Let:}
\[
\text{raw\_response\_txt} = \text{str}(\text{response\_obj})
\]
\textit{Then:}
\[
\text{passing} = \begin{cases} 
\text{True} & \text{if ``yes'' is in } \text{raw\_response\_txt.lower()} \\
\text{False} & \text{otherwise}
\end{cases}
\]
\[
\text{score} = \begin{cases}
1.0 & \text{if } \text{passing} \\
0.0 & \text{otherwise}
\end{cases}
\]
\textit{The output is:}
\[
\text{EvaluationResult} = \{q, r, \text{passing}, \text{score}, \text{feedback} = \text{raw\_response\_txt}, \text{contexts} = \{c_1, \dots, c_n\}\}
\]

\subsubsection{Context Relevancy}

\textit{Let} \( q \) be the query, \( \{c_1, c_2, \dots, c_n\} \) be the contexts. Define:

\[
\text{Documents} = \{d_i \mid d_i = \text{Document}(\text{text} = c_i)\}
\]\textit{}
\[
\text{Index} = \text{SummaryIndex}(\text{Documents})
\]
\textit{Evaluate the query} \( q \) \textit{using}:
\[
\text{query\_engine} = \text{Index.as\_query\_engine}(\text{llm}, \text{eval\_template}, \text{refine\_template})
\]
\[
\text{response\_obj} = \text{query\_engine.aquery}(q)
\]
\textit{Let:}
\[
\text{raw\_response\_txt} = \text{str}(\text{response\_obj})
\]
\textit{Parse the result:}
\[
\text{score}, \text{reasoning} = \text{parser\_function}(\text{raw\_response\_txt})
\]
\textit{Score threshold:}
\[
\text{score\_threshold} = 4.0
\]
\textit{Calculate:}
\[
\text{score} = \frac{\text{score}}{\text{score\_threshold}}
\]
\textit{Return:}
\[
\text{EvaluationResult} = \{q, \{c_1, \dots, c_n\}, \text{score}, \text{feedback} = \text{raw\_response\_txt}, \text{invalid\_result}, \text{invalid\_reason}\}
\]

\subsubsection{Analysis of Variance (ANOVA)}
\label{anova}
ANOVA is a fundamental statistical method used to compare means across multiple groups to determine if there are statistically significant differences between them. This study utilizes a one-way ANOVA, which examines the effect of a single independent variable - in this case, the type of RAG pipeline - on a dependent variable (performance metrics). 
The mean score reflects the average performance of each method across all 10 queries, offering an overall assessment of its effectiveness for the given metrics. A mean score closer to the highest possible value suggests that the method consistently delivers superior results, indicating strong performance across various queries. Conversely, a lower mean score points to weaker overall performance, highlighting areas where the method may be less effective. Essentially, the mean score serves as a summary indicator of each method's typical efficacy, providing a clear comparison of their relative strengths in achieving the desired outcomes.

The F-statistic in ANOVA quantifies the ratio of variance between groups to variance within groups, with larger F-values indicating greater differences among the groups. The degrees of freedom (\text{df}) are denoted as \text{F(2, 24)} for correctness and \text{F(2, 27)} for faithfulness and relevancy, indicating the number of independent values that can vary in the calculation. Here, the first value \text{(2)} represents the degrees of freedom for the groups (number of groups minus \text{1}), and the second value \text{(24/27)} represents the degrees of freedom for the error term (total observations minus the number of groups). The p-value assesses the probability of observing such differences between groups. Typically, a p-value below the significance level ($\alpha = \text{0.05}$) suggests statistically significant differences between the groups.

\subsection{Examples of Different Metrics}
\label{response}
\begin{figure}[H]
    \centering
    \includegraphics[width=30cm, height=8cm, keepaspectratio]{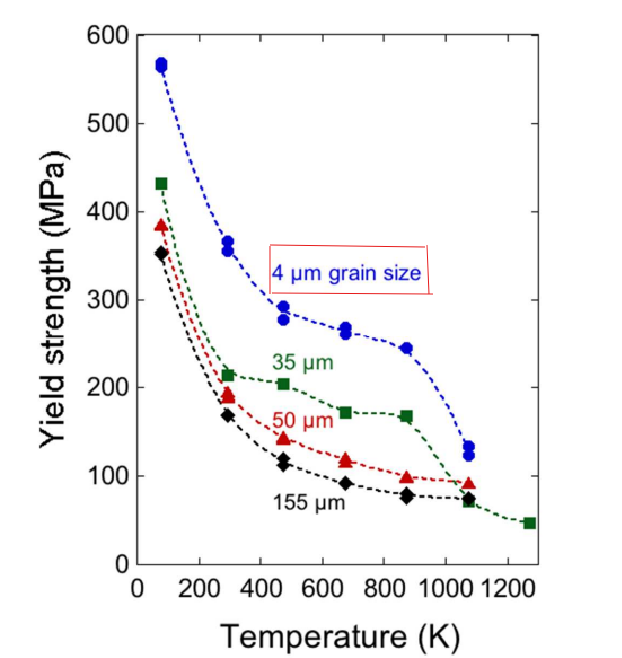}
    \caption{Information Extraction from Graph Images}
    \label{fig:graph}
\end{figure}

\begin{figure}[H]
    \centering
    \includegraphics[width=30cm, height=10cm, keepaspectratio]{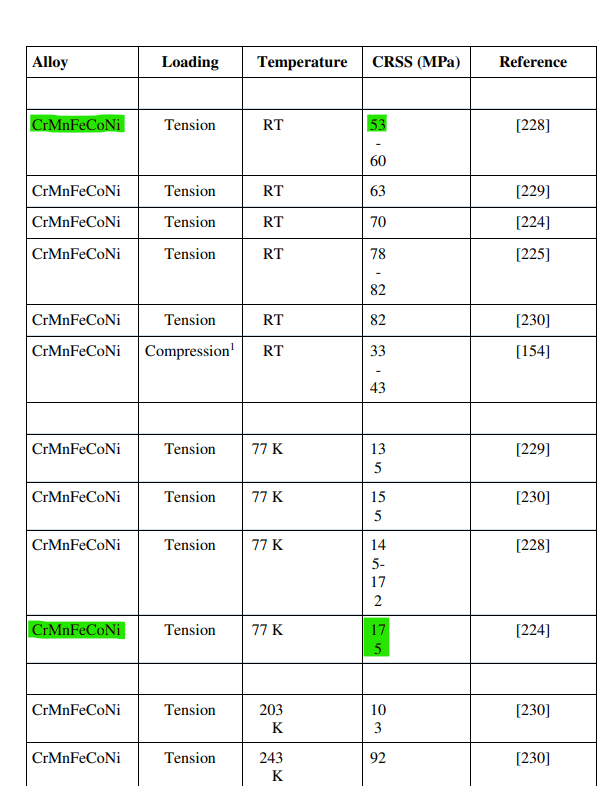}
    \caption{Information Extraction from Table Images}
    \label{fig:info_extraction_table}
\end{figure}

This section presents three example queries along with their corresponding responses from the RAG systems. Providing optimal responses to these queries requires effective information retrieval from text, figures, and tables. Additionally, we illustrate the application of various RAG evaluation metrics through these examples. Table \ref{context_relevancy} compares the performance using high-entropy alloy documents, demonstrating that G-RAG delivers contextually more relevant responses. The query in Table \ref{graph_query} necessitates accurate information extraction from Figure \ref{fig:graph}. The responses show that G-RAG with the parsing method outperforms other RAG systems, providing results that are closely aligned with the ground truth values. Finally, the query in Table \ref{table_query} requires accurate value extraction from the table presented in Figure \ref{fig:info_extraction_table}. In this case, G-RAG with the parsing method also outperforms other RAG systems. These results demonstrate that the integration of an efficient parsing method significantly enhances the performance of the G-RAG system.

\begin{table}[h!]
    \centering
    \caption{Comparison using \textbf{High-entropy alloy} Documents \textbf{(Context Relevancy)}}
    \renewcommand{\arraystretch}{1.75}
    \vspace{1em}
    \begin{tabular}{p{5cm} p{8.5cm}}
        \toprule
        \textbf{Query} & \textbf{What is Chromium?} \\
        \midrule
        G-RAG & Chromium is often a significant component in High-Entropy alloys (HEAs), particularly in the CoCrFeNi alloy, where it enhances corrosion resistance in both NaCl and H$_2$SO$_4$ solutions. However, when Chromium is replaced with Manganese in the Co-Cr-Fe-Ni alloy, the corrosion resistance in H$_2$SO$_4$ solution decreases, becoming inferior to 304 SS. The addition of certain elements like Al, Cu, B, and Mo can alter the corrosion resistance of various HEAs, often due to the formation of new phases or the impact on the existing phases' composition. The relationship between the degree of substitution ($\delta$
        ) and the enthalpy of mixing (H\textsubscript{mix}) for intermetallic compounds is an area of interest, potentially implying that as composition increases, the enthalpy of mixing decreases, which could have implications for the thermodynamics of alloys and intermetallic compound formation. \\
        Naive RAG & Chromium is a chemical element with the symbol Cr and atomic number 24. It is a hard, silver-white, and highly corrosion-resistant metal that is used in a variety of applications, including stainless steel, chrome plating, and catalytic converters in vehicles. \\
        Graph RAG & Chromium is a chemical element with the symbol Cr and atomic number 24. It is a steely-grey, lustrous, hard, and brittle transition metal. Chromium is also the main additive in stainless steel, to which it adds anti-corrosive properties. It is highly valued as a metal that is able to be highly polished while resisting tarnishing. Chromium is also used in various industrial applications, including the production of stainless steel, and chrome plating, and as a corrosion-resistant coating. \\
        \bottomrule
    \end{tabular}
    \label{context_relevancy}
    \vspace{1em} 
\end{table}

\newpage

\begin{table}[h!]
    \centering
    \caption{Comparison of Correct Information Extraction from Graph}
    \renewcommand{\arraystretch}{1.75} 
    \vspace{1em}
    \begin{tabular}{p{5cm} p{8cm}}
        \toprule
        \textbf{Query} & \textbf{What is the yield strength of the CrMnFeCoNi alloy at 600 K, 700 K with 4 µm grain size?} \\
        \midrule
        \textbf{Ground Truth} & The yield strength of the CrMnFeCoNi alloy at 600 K is 290 MPa, and at 700 K, it is 285 MPa. \\
        (G-RAG + \textbf{Parsing Method})  & The yield strength of CrMnFeCoNi alloy at 600 K is 310 MPa, and at 700 K is around 300 MPa. \\
        Naive RAG & Unfortunately, the information provided does not contain any data about the yield strength of the CrMnFeCoNi alloy at 600 K, 700 K, or with a 4 µm grain size. \\
        G-RAG & The yield strength of CrMnFeCoNi alloy at 600 K is 450 MPa, and at 700 K is 400 MPa. \\
        \bottomrule
    \end{tabular}
    \label{graph_query}
    \renewcommand{\arraystretch}{1} 
\end{table}

\newpage

\begin{table}[h!]
    \centering
    \caption{Comparison of Correct Information Extraction from Table}
    \renewcommand{\arraystretch}{1.75} 
    \vspace{1em}
    \begin{tabular}{p{5cm} p{8cm}}
        \toprule
        \textbf{Query} & \textbf{What is the CRSS of CrMnFeCoNi at the tension in room temperature?} \\
        \midrule
        \textbf{Ground Truth} & The Critical Resolved Shear Stress (CRSS) of the CrMnFeCoNi alloy has been measured at 53 MPa at room temperature and 175 MPa at 77 K. \\
        (G-RAG + \textbf{Parsing Method})  &  The CRSS of the CrMnFeCoNi alloy has been determined to be 53 MPa at room temperature and 175 MPa at 77 K. \\
        Naive RAG & Unfortunately, the provided context does not contain information about the CRSS (Critical Resolved Shear Stress) of CrMnFeCoNi at tension at room temperature. \\
        G-RAG & Sorry, I do not have any information about the CRSS of  CrMnFeCoNi at the tension at room temperature.  \\
        \bottomrule
    \end{tabular}
    \label{table_query}
    \renewcommand{\arraystretch}{1} 
\end{table}

\begin{thebibliography}{00}
\bibitem{gaoa2023retrieval} Yunfan Gaoa, Y Xiong, X Gao, K Jia, J Pan, Y Bi, Y Dai, J Sun, M Wang, and Haofen
 Wang. Retrieval-augmented generation for large language models: A survey. \textit{arXiv preprint
 arXiv:2312.10997,} 2023.
\bibitem{sharma2024retrieval} Sanat Sharma, David Seunghyun Yoon, Franck Dernoncourt, Dewang Sultania, Karishma
 Bagga, Mengjiao Zhang, Trung Bui, and Varun Kotte. Retrieval augmented generation for
 domain-specific question answering. \textit{arXiv preprint arXiv:2404.14760}, 2024.

\bibitem{ma2024think} Shengjie Ma, Chengjin Xu, Xuhui Jiang, Muzhi Li, Huaren Qu, and Jian Guo. Think-on-graph
 2.0: Deep and interpretable large language model reasoning with knowledge graph-guided
 retrieval. \textit{arXiv e-prints,} pages arXiv–2407, 2024.

\bibitem{sandhu2024exploration}  Tanvi Sandhu. Exploration of word embeddings with graph-based context adaptation for
 enhanced word vectors. Master’s thesis, University of Windsor (Canada), 2024.

\bibitem{edge2024local}  Darren Edge, Ha Trinh, Newman Cheng, Joshua Bradley, Alex Chao, Apurva Mody, Steven
 Truitt, and Jonathan Larson. From local to global: A graph rag approach to query-focused
 summarization. \textit{arXiv preprint arXiv:2404.16130,} 2024.

\bibitem{park2024leveraging}  Chaelim Park, Hayoung Lee, and Ok-ran Jeong. Leveraging medical knowledge graphs and
 large language models for enhanced mental disorder information extraction. \textit{Future Internet,}
 16(8):260, 2024.

\bibitem{buehler2024generative} Markus J Buehler. Generative retrieval-augmented ontologic graph and multiagent strategies for
 interpretive large language model-based materials design. \textit{ACS Engineering Au,} 4(2):241–277,
 2024.

\bibitem{elile2024graph} Julien Delile, Srayanta Mukherjee, Anton Van Pamel, and Leonid Zhukov. Graph-based retriever
 captures the long tail of biomedical knowledge. \textit{arXiv preprint arXiv:2402.12352,} 2024.

\bibitem{peng2023knowledge} Ciyuan Peng, Feng Xia, Mehdi Naseriparsa, and Francesco Osborne. Knowledge graphs:
 Opportunities and challenges. \textit{Artificial Intelligence Review,} 56(11):13071–13102, 2023.

\bibitem{anokhin2024arigraph} Petr Anokhin, Nikita Semenov, Artyom Sorokin, Dmitry Evseev, Mikhail Burtsev, and Evgeny
 Burnaev. Arigraph: Learning knowledge graph world models with episodic memory for llm
 agents. \textit{arXiv preprint arXiv:2407.04363,} 2024.

\bibitem{merth2024superposition} ThomasMerth, Qichen Fu, MohammadRastegari, and Mahyar Najibi. Superposition prompting:
 Improving and accelerating retrieval-augmented generation. \textit{arXiv preprint arXiv:2404.06910,}
 2024.

\bibitem{chen2023extending} Shouyuan Chen, Sherman Wong, Liangjian Chen, and Yuandong Tian. Extending context
 window of large language models via positional interpolation. \textit{arXiv preprint arXiv:2306.15595,}
 2023.

\bibitem{hu2024memserve} Cunchen Hu, Heyang Huang, Junhao Hu, Jiang Xu, Xusheng Chen, Tao Xie, Chenxi Wang,
 Sa Wang, Yungang Bao, Ninghui Sun, et al. Memserve: Context caching for disaggregated llm
 serving with elastic memory pool. \textit{arXiv preprint arXiv:2406.17565,} 2024.

\bibitem{wang2024searching}  Xiaohua Wang, Zhenghua Wang, Xuan Gao, Feiran Zhang, Yixin Wu, Zhibo Xu, Tianyuan
 Shi, Zhengyuan Wang, Shizheng Li, Qi Qian, et al. Searching for best practices in retrieval
augmented generation. In \textit{Proceedings of the 2024 Conference on Empirical Methods in Natural
 Language Processing,} pages 17716–17736, 2024.

\bibitem{liu2024lost}  Nelson F Liu, Kevin Lin, John Hewitt, Ashwin Paranjape, Michele Bevilacqua, Fabio Petroni,
 and Percy Liang. Lost in the middle: How language models use long contexts. \textit{Transactions of
 the Association for Computational Linguistics,} 12:157–173, 2024.

\bibitem{peng2024graph} Boci Peng, Yun Zhu, Yongchao Liu, Xiaohe Bo, Haizhou Shi, Chuntao Hong, Yan Zhang,
 and Siliang Tang. Graph retrieval-augmented generation: A survey. \textit{arXiv preprint
 arXiv:2408.08921,} 2024.

\bibitem{tedeschi2021named} Simone Tedeschi, Simone Conia, Francesco Cecconi, and Roberto Navigli. Named entity
 recognition for entity linking: What works and what’s next. In \textit{Findings of the Association for
 Computational Linguistics: EMNLP 2021,} pages 2584–2596, 2021.

\bibitem{orlando2024relik} Riccardo Orlando, Pere-Lluís Huguet Cabot, Edoardo Barba, and Roberto Navigli. Relik:
 Retrieve and link, fast and accurate entity linking and relation extraction on an academic budget.
 \textit{arXiv preprint arXiv:2408.00103,} 2024.

\end{thebibliography}
\end{document}